# Making sense of global collaboration dynamics: Developing a methodological framework to study (dis)similarities between country disciplinary profiles and choice of collaboration partners

Nicolas Robinson-Garcia[1], Richard Woolley[2] and Rodrigo Costas[3]

*[1] elrobinster@gmail.com*
Delft Institute of Applied Mathematics, TU Delft, Netherlands
INGENIO (CSIC-UPV), Universitat Politècnica de València, Spain

*[2] ricwoo@ingenio.upv.es*
INGENIO (CSIC-UPV), Universitat Politècnica de València, Spain

*[3] rcostas@cwts.leidenuniv.nl*
CWTS, Leiden University, Netherlands
DST-NRF Centre of Excellence in Scientometrics and Science, Technology and Innovation Policy, Stellenbosch University, South Africa

**Abstract**
This paper presents a novel methodological framework by which the effects of globalization on international collaboration can be studied and understood. Using the cosine similarity of the disciplinary and partner profiles of countries by collaboration types it is possible to analyse the effects of globalization and the costs and benefits of an increasing global networked research system.

**Introduction**

The growth of scientific collaboration and co-authorship in the last century has drawn the attention of bibliometricians and sociologists of science for quite some time now. In the words of Cole (1973), "it is generally recognized today that science in fact develops within a community of interacting scientists" (p. 1). The seminal works by Beaver & Rosen (1978) and Crane (1972) have led to a profusion of studies analysing collaboration in many ways, including motivations for collaboration (Katz & Martin, 1997), collaboration strategies (Bozeman & Corley, 2004), collaboration structure (Newman, 2001) or benefits derived from international collaboration in terms of scientific impact (Bote, Olmeda-Gómez, & Moya-Anegón, 2013), among others.

The increase of international collaboration experienced in the last decades has been described as an effect of the globalization of the research system, leading to an increase of international exchange and flows of mobile scholars (Czaika & Orazbayev, 2018), and facilitating collaboration between distant countries (Waltman, Tijssen, & Eck, 2011). It has been argued that such changes could decrease national disparities by 'flattening' the world and reducing zones of inclusion and exclusion in a global scientific collaboration network (Saxenian, 2005; Woolley, Robinson-Garcia, & Costas, 2017). But empirical studies suggest that international collaboration networks are organized hierarchically (Wagner, Whetsell, & Leydesdorff, 2017), showing an inverse relation between the growth of the share of collaborative papers and the geographical spread of these networks (Wagner, 2005). These findings along with the fact that networked or multilateral collaboration is increasing overtime (see table 1) raise many questions with regard to how countries are adapting to this new landscape, their integration into these global networks and the potential 'costs' or 'benefits' this integration may bring together.

A fundamental element of understanding of the dynamics of integration in global scientific networks is based on the work of the German sociologist Georg Simmel (1950: 135), who famously recognised the effect of introducing a third actor into a social context: "there is, in addition to the direct relation between A and B, for instance, their indirect one, which is derived from their common relation to C". Simmel took this expansion from the dyad to the triad as the basis for a crucial piece of understanding, that indirect relationships are essential to the formation and cohesion of the groups and sub-groups that characterize the social world. Social network analysis has since built on this foundational distinction to investigate social structures (Burt 1992) and study how the roles and positioning of actors in social networks effects the distribution of power (Brass and Burkhardt 1992) and the diffusion of information (Granovetter 1973).

We take up this distinction between dyads and triads as the basis for our understanding of two modes of international collaboration in science: partnerships and networks. International research partnerships and networks are both constructed from the same fundamental element, what network theorists call "mutual dyads" in which the relationship between two actors is based on mutual recognition and reciprocity (Wasserman and Faust 1994: 511). We only focus on mutual dyads (and not on 'directional' dyads) because we tend to assume that collaboration relationships in science are based on direct interpersonal relationships characterized by varying degrees of both trust and conflict (Shrum et al. 2001). However, whereas we understand international research partnerships to be based on a single mutual dyad, we understand international research networks as being based on a combination of two or more mutual dyads. This network mode can be illustrated most simply in a triadic form (Figure 0).

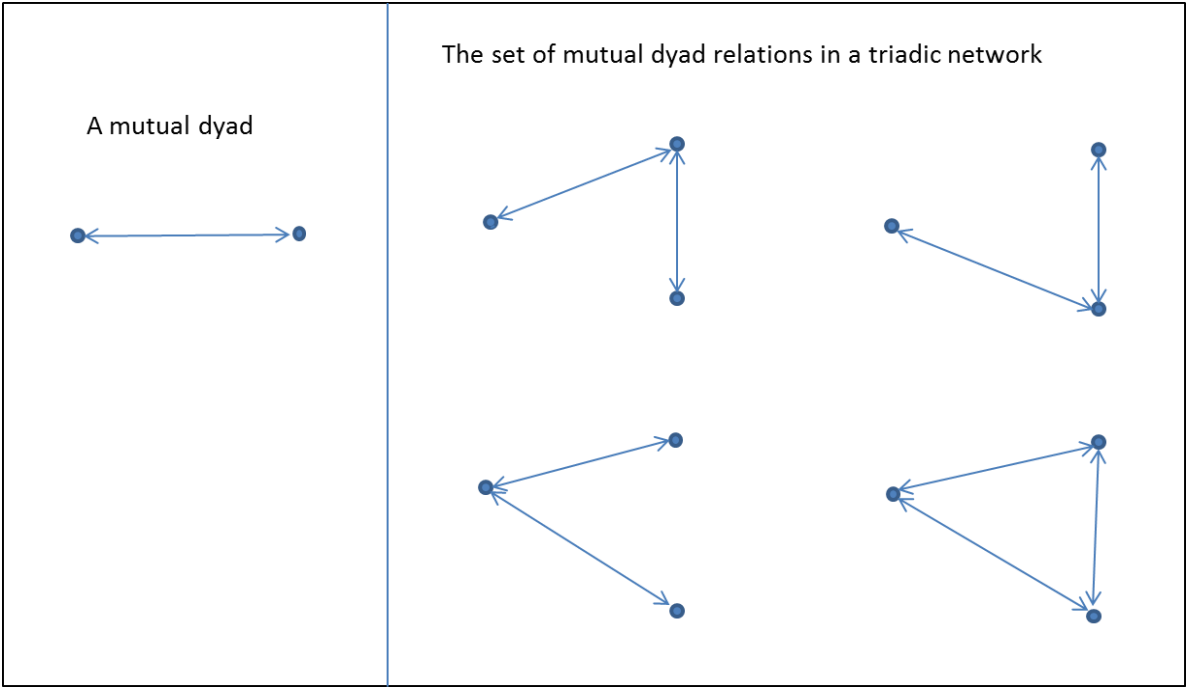

**Figure 0. Mutual dyads in triadic relationships**

A triadic collaboration network can be made up of four potential sets of mutual dyads (Fig 1). In three of the four possible triadic networks the relationship between two of the nodes only *only* indirect, that is, mediated via the third node. International research collaborations can thus be characterised theoretically as containing "structural holes" (Burt 1992), where the triangular relationship is not closed on all sides by a mutual dyad. We apply this logic to international scientific collaboration to assume that 1) whilst all collaboration relationships are based on mutual reciprocity and exchange based on trust relations; 2) it is not necessarily the case that all collaborators must have a direct relationship with all other collaborators. While network theorists argue that "transitive" networks, in which the triad is 'closed' and all elements are connected are more stable and likely to be more durable (Burt 1992; Wasserman and Faust 1994), we simply do not know whether this is the case in relation to scientific collaborations or not. Rather, a very significant limitation of network analyses based on bibliometric data is that the qualities of networks remain obscured.

The distinction between bilateral international research collaboration (BIRC) and multilateral international research collaboration (MIRC) is thus important because discussions of global co-authorship networks implicitly assume equality of connectedness within the network, assuming that a homogeneous set of mutual dyadic relations exist between each co-author of a scientific paper. In other words, a default assumption of homogeneity in sets of bibliometric relationships slides into an assumption of equality in sets of sociological relationships. While this is not intended as a criticism of what bibliometric-based network analyses can do, it is intended as a reminder of the limitations of bibliometric network analyses in relation to assuming 'connectedness' as a homogeneous form of social organisation of knowledge production - and then characterising so-called 'global networks' on this basis. Although we are not able to understand whether the multilateral networks in our data are transitive (closed) or not, this theoretical approach will nevertheless have significant implications for how we interpret our results.

**Table 1. Growth rate by WorldBank regions for the 1980-2017 period for Bilateral International Research Collaboration (BIRC) and Multilateral International Research Collaboration (MIRC) collaboration. Data source: Web of Science**

|  | BIRC | MIRC |
|---|---|---|
| **East Asia & Pacific** | 11.4% | 15.8% |
| **Europe & Central Asia** | 7.4% | 12.6% |
| **Latin America & Caribbean** | 9.5% | 15.2% |
| **Middle East & North Africa** | 8.8% | 14.3% |
| **North America** | 7.4% | 12.6% |
| **South Asia** | 9.8% | 16.1% |
| **Sub-Saharan Africa** | 8.3% | 13.3% |

In this study, we propose a new methodological framework by which the dynamics of globalization through international scientific collaboration can be studied and start to be better understood. For this, we suggest comparing countries' disciplinary profiles and choice of collaboration partner (i.e. the collaboration partner profile) by publication type. We distinguish the following publication types: 1) domestic publication (publications authored by one or more scholars affiliated to a single country), 2) bilateral international collaboration (publications co-authored by scholars affiliated to two different countries) and 3) participating in a collaboration network (publications co-authored by scholars affiliated to three or more countries). We base our empirical work on the model proposed by (De Lange & Glänzel, 1997; Glänzel & De Lange, 1997) who previously distinguished between no collaboration, BIRC and MIRC. We hypothesize that countries' international profile will differ from their domestic profile, with this difference being greater for their MIRC profile. In other words, as countries become drawn into multilateral international networks they move away from the focus (topics) that characterise domestic knowledge production. We suggest that this methodology has the potential to better inform studies focused on specific countries or regions to better understand not only how integrated they are in global networks, but also the potential effects or factors which can explain their specific situations.

**Data and methods**

*Data collection*

We collected all publications for all countries for the 2008-2017 period. This data was gathered from the CWTS-enhanced version of the Web of Science. Country information was extracted from each publication, normalized and linked to the World Bank regions classification. For each publication we counted the number of different countries to which authors were affiliated and created a new collaboration type field in which we distinguished between BIRC, MIRC and domestic publications. Disciplinary profiles of countries are created based on the distribution of their publications among the Web of Science subject categories classification. We produced four disciplinary profiles for each country: 1) based on their domestic output, 2) based on their international output (co-authored with at least another country), 3) based on their BIRC output (co-authored just with one other country) and 4) based on their MIRC output (co-authored with at least two other countries). This distinction is based on the model used by (De Lange & Glänzel, 1997; Glänzel & De Lange, 1997), although we acknowledge that this approach could be extended by including further divisions. For instance, Adams & Gurney (2018) suggest that publications authored by 20 or more countries should be treated differently due to their special nature.

*Methodological approach*

Here we propose measuring similarity of a countries' domestic disciplinary profile with their BIRC and MIRC disciplinary profiles by using the cosine similarity (Salton & McGill, 1986). Cosine similarity is usually employed in bibliometric studies when analyzing co-occurrence data such as co-citation networks or co-citation networks (e.g., Aman, 2018; Wagner, 2005; Yan & Ding, 2012).

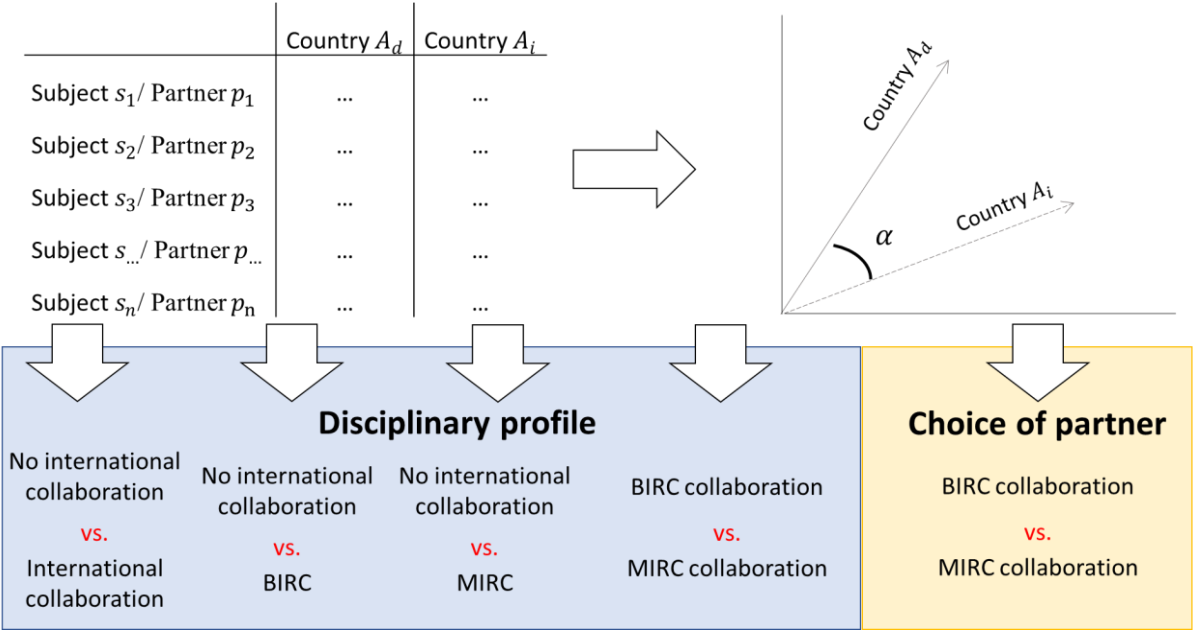

**Figure 1. Calculation of five similarity variables per country for their disciplinary and collaboration partners profiles.**

Figure 1 presents a schematic representation of our approach, depicting the profile distribution of subjects ($s_n$) or partners ($p_n$), for domestic publications ($A_{dom}$) and international publications ($A_{int}$), and how the cosine analysis is applied for the disciplinary profile on the one hand, and the choice of partner on the other hand. Let $A_d$ be the disciplinary profile of country A when publishing domestically where $A_{dom} = \{s_1, s_2, \ldots, s_n\}$ being $s_n$ the number of publications $n$ in subject $s$. Let $A_{int}$ be the disciplinary profile of country A when publishing with international collaboration. The similarity between two disciplinary profiles is defined as:

$$SIM(A_{dom}, A_{int}) = \cos(\alpha) = \frac{A_{dom}\, A_{int}}{|A_{dom}||A_{int}|}$$

Where a value of 0 indicates no similarity between profiles and 1 indicates that both profiles are identical. This same procedure can then be applied to all combinations of publication types to find (dis)similarities between disciplinary profiles. Furthermore, it can also be calculated to identify (dis)similarities on countries' partner of choice distribution. To this end, we compute five similarity indicators for each country as shown in Figure 1. While four of them relate to disciplinary similarity, one relates to similarity on collaboration partner. The rationale of this is that one would expect that a high disciplinary (dis)similarity between BIRC and MIRC profiles, would lead to a high (dis)similarity in the distribution of publications by collaboration partners between BIRC and MIRC profiles.

## Results

In this section we show some results of a global analysis of international collaboration. Figure 2 shows the proportion of BIRC for each region in the world. In general, the majority of regions' output in international collaboration is bilateral with the exception of Sub-Saharan Africa (on median, 46% of their output is bilateral). On the other extreme we find North America (69%) or South Asia (64%). Overall we find an important dispersion among countries within all regions, with no significant differences between regions (with the exception of Sub-Saharan Africa and North America.

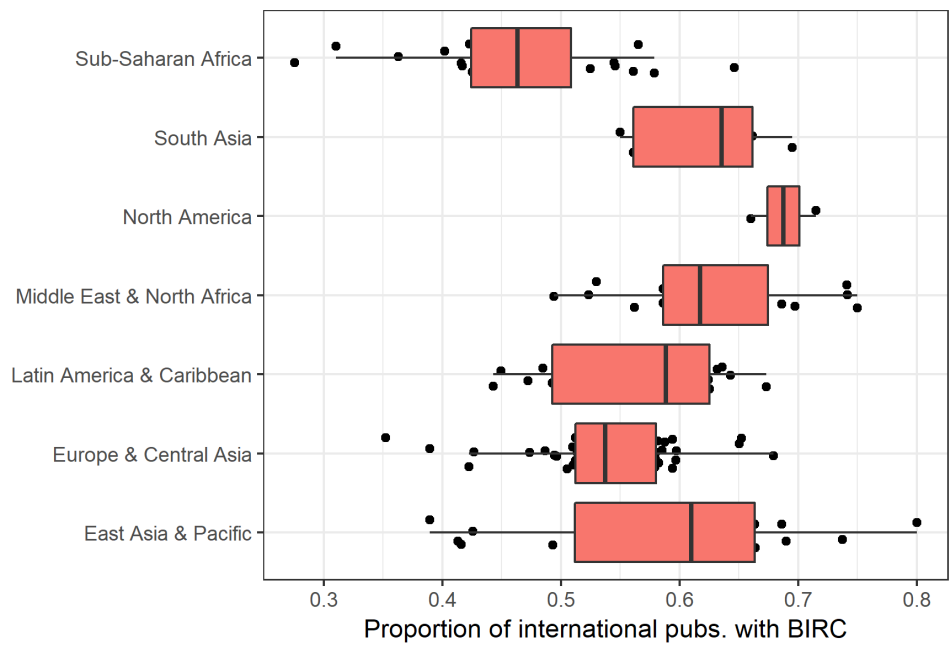

**Figure 2. Proportion of publications with BIRC by WorldBank regions for the 2008-2017 period**

Figure 3 illustrates how cosine similarities between the different profiles of a country disaggregated by publication type, can be used to better understand disciplinary differences between countries. The figure highlights similarities and differences between international collaboration and domestic knowledge production (Figure 3A), between collaborating bilaterally and collaborating multilaterally (Figure 3B), and between choice of partner countries when collaborating bilaterally and collaborating multilaterally (Figure 3C). In these figures, countries in blue are those which exhibit a similarity of profiles above the world average, while those in orange exhibit a similarity below world average. Overall, we observe that differences are more acute between the disciplinary profiles of countries when collaborating internationally versus not collaborating than they are between BIRC and MIRC. Furthermore, we observe that, with the exception of the United States on choice of partner (Figure 3C), countries with a similarity below average tend to belong to Eastern Europe, Africa and, to some extent, Asia.

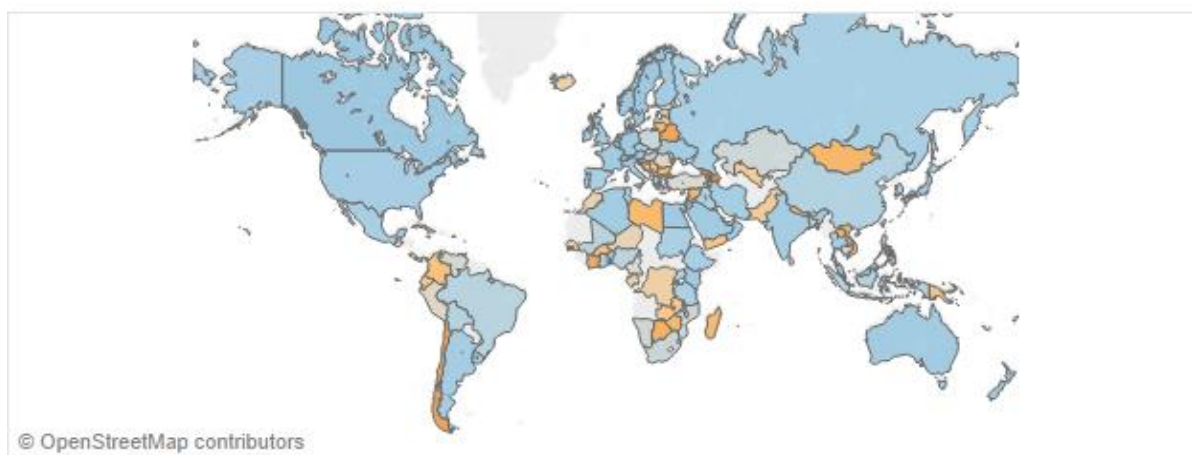

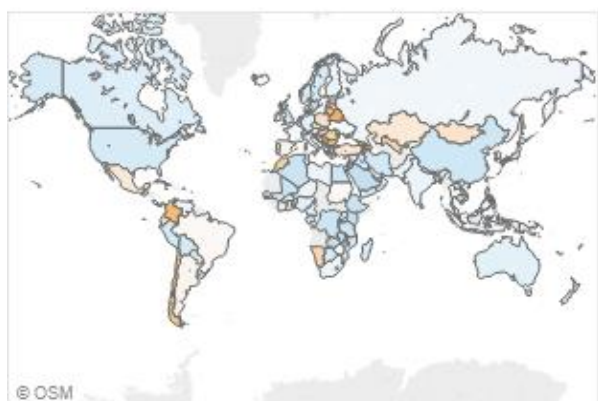

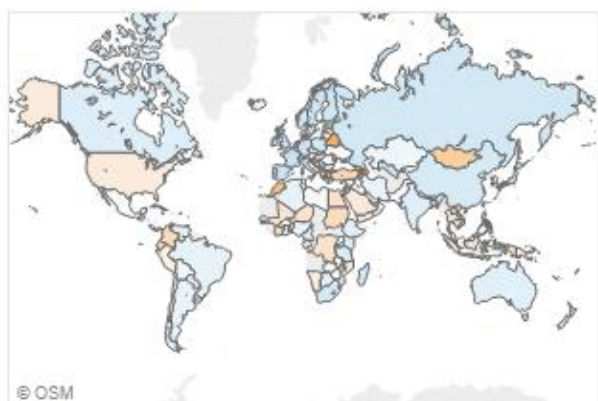

**Figure 3. Cosine similarities for each country according to their A) disciplinary domestic and international profiles, B) disciplinary BIRC and MIRC profiles, and C) collaboration partners' BIRC and MIRC profiles. Blue intensity indicates similarity above world average; Orange intensity indicates similarity below world average.**

When looking into which type of international collaboration deviates more from the disciplinary profile of countries' domestic publications (Figure 4), we observe that MIRC does appear to be more dissimilar than BIRC profiles. But we can also see that these differences vary greatly between countries and regions. For instance, we observe large differences for Latin America & Caribbean, and specifically for Chile and Colombia. Also Eastern European and Central Asian countries seem to show a larger difference than Western and Central European countries. Indeed, we observe that while the similarity of European and Central Asian countries between their BIRC and national disciplinary profiles is high, it is also relatively homogeneous between countries, while there are larger disparities when comparing the similarity of their MIRC and national disciplinary profiles.

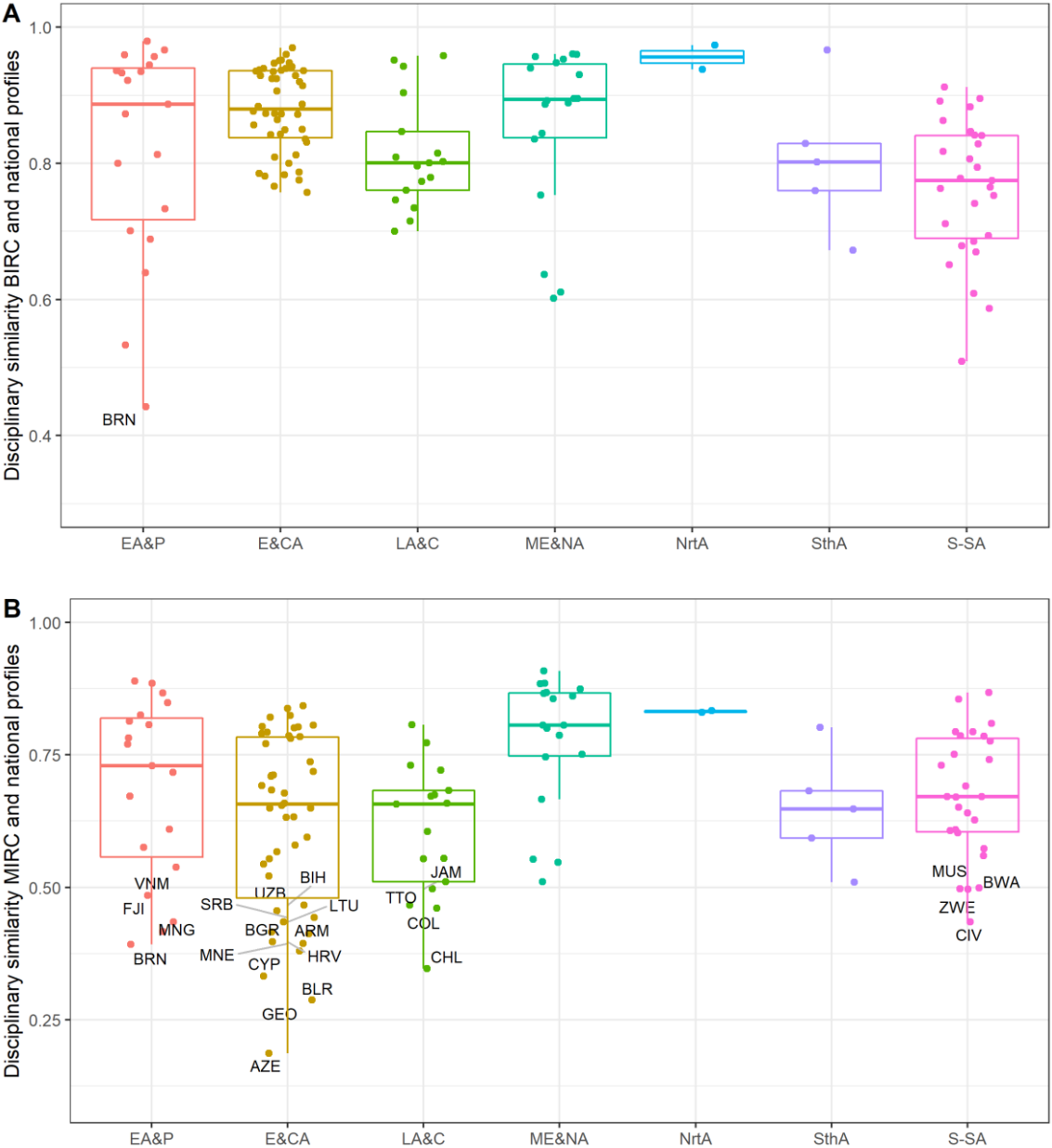

**Figure 4. Boxplots showing similarity of countries by region for A) their BIRC versus national and B) their MIRC versus national disciplinary profile. Countries with value under 0.5 are highlighted**

Figure 5 and 6 illustrate further ways by which these indicators could be employed to better interpret specific situations. Figure 5 shows in the y-axis the similarity of the disciplinary profile of countries from East Asia & Pacific when collaborating internationally versus when not doing so (domestic profile). The x-axis shows the proportion of publications internationally co-authored they produce. We observe that the least productive countries are not only the ones which show a higher dependency on international collaboration, but many of these smaller countries exhibit greater disciplinary differences when collaborating internationally from the domestic profile.

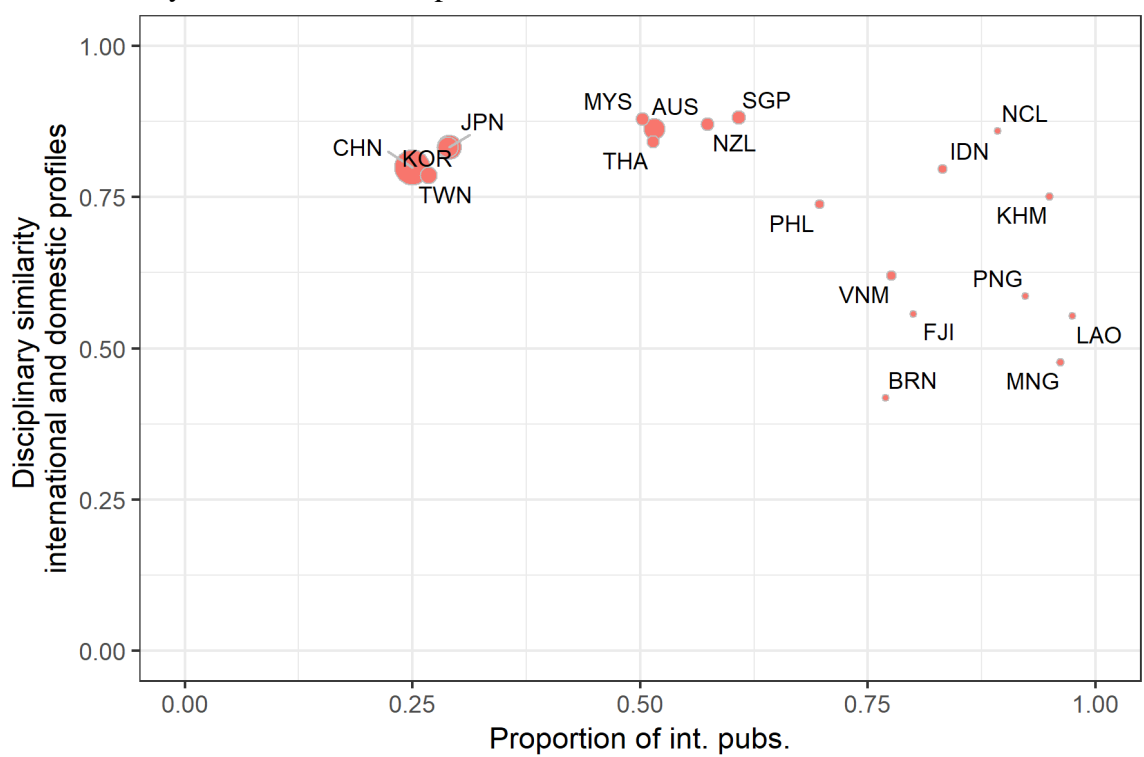

**Figure 5. Scatterplot comparing the disciplinary similarity of the international versus domestic profile of each country and the proportion of publications internationally co-authored for countries in East Asia & Pacific. Size of dots indicates total number of publications**

In Figure 6 we look into both Europe & Central Asia and Sub-Saharan Africa and compare similarities between countries' BIRC and MIRC disciplinary profiles with their BIRC and MIRC choice of partner. Countries that exhibit lower disciplinary similarities also exhibit lower similarities on choice of partners in Europe and Central, however this is not always the case for Sub-Saharan African countries. Western European countries show a higher similarity for both indicators, while interestingly South Africa shows a high partner similarity but a lower level of disciplinary similarity when comparing BIRC and MIRC

**Concluding remarks**

In this paper we propose deconstructing countries' publication profile based on the collaboration type of their output. We suggest that by using similarity measures and comparing countries' collaboration profiles both in terms of their distribution of disciplines and choice of partner, we take an important step toward the development of a better understanding of the international partnerships and networks that shape the dynamics of globalising science.

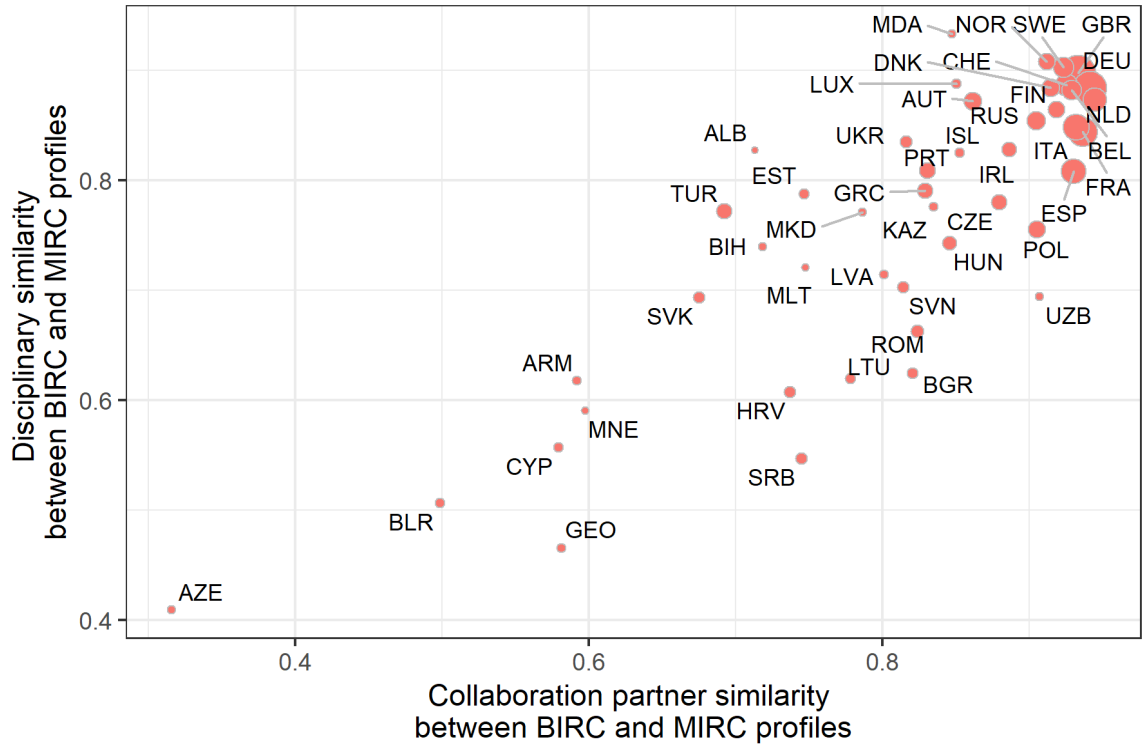

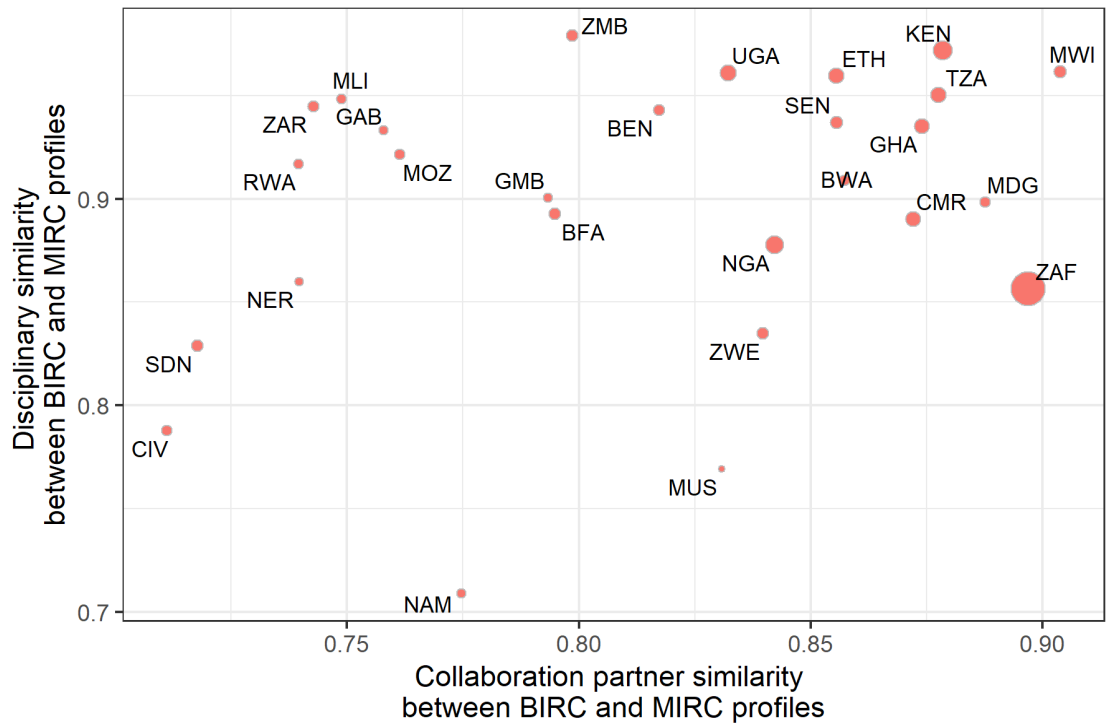

**Figure 6. Scatterplot comparing the disciplinary similarity of the BIRC versus MIRC and the collaboration partner's similarity of the BIRC versus MIRC profile for each country in Europe & Central Asia (above) and Sub-Saharan Africa (below). Size of dots indicates number of internationally co-authored publications**

We have suggested the use of the cosine as a measure of similarity to establish a set of 'internal' comparisons at the national level, in the sense that we are always comparing different portions of a country's output. We have shown how this method can help us to understand the connections between what a country's scientists are working on (topic, discipline) and who they are working on it with (international partners). We have also shown the efficacy of the method for comparing between countries and regions on these dimensions. Further methodological development will follow to identify other types of benchmarking of international collaboration that can be designed from this foundation.

Our theoretical distinction between international research partnerships (BIRC) and international networks (MIRC) can also form the basis for future efforts to better understand relationships between formalised patterns of scientific co-authorship and the social integration underpinning scientific collaborations. Further technical advance and the use of complementary methodologies will be required in this respect. But we consider the present methodological innovation to be a constructive first step in this direction.

## Acknowledgments

Nicolas Robinson-Garcia is a Marie Sklodowska Curie experienced researcher in the LEaDing Fellows COFUND programme sponsored by the European Commission. This research is partially funded by the South African DST-NRF Centre of Excellence in Scientometrics and Science, Technology and Innovation Policy (SciSTIP).

## References

Adams, J., & Gurney, K. A. (2018). Bilateral and Multilateral Coauthorship and Citation Impact: Patterns in UK and US International Collaboration. *Frontiers in Research Metrics and Analytics*, *3*. Recuperado mayo 4, 2018, a partir de https://www.frontiersin.org/articles/10.3389/frma.2018.00012/full

Aman, V. (2018). A new bibliometric approach to measure knowledge transfer of internationally mobile scientists. *Scientometrics*, *117*(1), 227-247.

Beaver, D. deB, & Rosen, R. (1978). Studies in scientific collaboration. *Scientometrics*, *1*(1), 65-84.

Bote, V. P. G., Olmeda-Gómez, C., & Moya-Anegón, F. de. (2013). Quantifying the benefits of international scientific collaboration. *Journal of the American Society for Information Science and Technology*, *64*(2), 392-404.

Bozeman, B., & Corley, E. (2004). Scientists' collaboration strategies: implications for scientific and technical human capital. *Research Policy*, Scientific and Technical Human Capital: Science Careers and Networks as Knowledge Assets, *33*(4), 599-616.

Brass D.J. & Burkhardt, M.E. (1992). Centrality and power in organizations. In: Nohria N and Eccles RG (eds) *Networks and Organizations: Structure, Form, and Action*. Cambridge, MA: Harvard Business School Press, pp. 191–215.

Burt, R. (1992). *Structural Holes: The Social Structure of Competition*. Cambridge, MA: Harvard University Press.

Chinchilla-Rodríguez, Z., Bu, Y., Robinson-García, N., Costas, R., & Sugimoto, C. R. (2018). Travel bans and scientific mobility: utility of asymmetry and affinity indexes to inform science policy. *Scientometrics*, 1-22.

Cole, J.R. (1973). *Social stratification in science.* Chicago ; London: University of Chicago Press.

Crane, D. (1972). *Invisible colleges: Diffusion of knowledge in scientific communities*. Chicago: University of Chicago Press.

Czaika, M., & Orazbayev, S. (2018). The globalisation of scientific mobility, 1970–2014. *Applied Geography*, *96*, 1-10.

De Lange, C., & Glänzel, W. (1997). Modelling and measuring multilateral co-authorship in international scientific collaboration. Part I. Development of a new model using a series expansion approach. *Scientometrics*, *40*(3), 593–604.

Glänzel, W., & De Lange, C. (1997). Modelling and measuring multilateral co-authorship in international scientific collaboration. Part II. A comparative study on the extent and change of international scientific collaboration links. *Scientometrics*, *40*(3), 605–626.

Granovetter MS (1973) The strength of weak ties. *American Journal of Sociology* 78(6): 1360–1380.

Katz, J. S., & Martin, B. R. (1997). What is research collaboration? *Research Policy*, *26*(1), 1-18.

Newman, M. E. J. (2001). The structure of scientific collaboration networks. *Proceedings of the National Academy of Sciences*, *98*(2), 404-409.

Salton, G., & McGill, M. J. (1986). Introduction to modern information retrieval.

Saxenian, A. (2005). From Brain Drain to Brain Circulation: Transnational Communities and Regional Upgrading in India and China. *Studies in Comparative International Development*, *40*(2), 35-61.

Shrum, W., Compalov, I:, & Genuth, J. (2001) Trust, Conflict and Performance in Scientific Collaborations. *Social Studies of Science* 31(5): 681-730. doi: 10.1177/030631201031005002

Simmel, G. (1950). *The Sociology of George Simmel*. K Wolf (Ed.), Glencoe, Illinois: Free Press.

Wagner, C. S. (2005). Six case studies of international collaboration in science. *Scientometrics*, *62*(1), 3-26.

Wagner, C. S., Whetsell, T. A., & Leydesdorff, L. (2017). Growth of international collaboration in science: revisiting six specialties. *Scientometrics*, *110*(3), 1633-1652.

Waltman, L., Tijssen, R. J. W., & Eck, N. J. van. (2011). Globalisation of science in kilometres. *Journal of Informetrics*, *5*(4), 574-582.

Wasserman, S. & Faust, K. (1994) *Social Network Analysis: Methods and Applications*. Cambridge: CUP:

Woolley, R., Robinson-Garcia, N., & Costas, R. (2017). Global research collaboration: Networks and partners in South East Asia. *arXiv:1712.06513 [cs]*. Recuperado marzo 28, 2018, a partir de http://arxiv.org/abs/1712.06513

Yan, E., & Ding, Y. (2012). Scholarly Network Similarities: How Bibliographic Coupling Networks, Citation Networks, Cocitation Networks, Topical Networks, Coauthorship Networks, and Coword Networks Relate to Each Other. *Journal of the American Society for Information Science and Technology*, *63*(7), 1313-1326.